\documentclass[twocolumn,showpacs,amssymb,10pt]{revtex4}
\usepackage{amssymb}
\usepackage{amsmath}
\usepackage{subfigure}
\usepackage{natbib}
\usepackage{amsfonts}
\usepackage{graphicx}
\usepackage{dcolumn}
\usepackage{bm}
\usepackage{epsfig}
\usepackage{float}
\usepackage{makeidx}
\usepackage{color}
\graphicspath{{figs/}}
\usepackage{longtable}

\begin{document}

\preprint{}

%
%

\title{The effects of KSEA interaction on the ground-state properties of spin chains in a transverse field}

%
%
\author{Hao Fu}
\affiliation{Department of Physics and Institute of Theoretical Physics, Nanjing Normal
University, Nanjing 210023, P. R. China}
\author{Ming Zhong}
\affiliation{Department of Physics and Institute of Theoretical Physics, Nanjing Normal
University, Nanjing 210023, P. R. China}
\author{Peiqing Tong}

\email{pqtong@njnu.edu.cn}

\affiliation{Department of Physics and Institute of Theoretical Physics, Nanjing Normal
University, Nanjing 210023, P. R. China}
\affiliation{Jiangsu Key Laboratory for Numerical Simulation of Large Scale Complex Systems,
Nanjing Normal University, Nanjing 210023, P. R. China}

\date{\today \\[0.3cm]}

%
%
\begin{abstract}
The effects of symmetric helical interaction which is called the Kaplan, Shekhtman, Entin-Wohlman, and Aharony (KSEA) interaction on the ground-state properties of three kinds of spin chains in a transverse field have been studied by means of correlation functions and chiral order parameter. We find that the anisotropic transition of $XY$ chain in a transverse field ($XY$TF) disappears because of the KSEA interaction.
For the other two chains, we find that the regions of gapless chiral phases in the parameter space induced by the DM or $XZY-YZX$ type of three-site interaction are decreased gradually with increase of the strength of KSEA interaction. When it is larger than the coefficient of DM or $XZY-YZX$ type of three-site interaction, the gapless chiral phases also disappear.
\end{abstract}
\maketitle

%
%

%
%
%
%
%
\section{Introduction}\label{sec:intoduc}
    In general, quantum phase transitions (QPTs) are phase transitions that the ground state of a quantum system undergoes a qualitative change by variation in a parameter of the system \cite{g7,yy15}.
    Unlike thermodynamic phase transitions, QPTs refer to the phase transitions of the system by quantum fluctuations at absolute zero temperature.
    The QPTs have been extensively studied in the quantum many-body system \cite{yy1,yy2,yy3,yy4,yy5} and strongly correlated system \cite{yy6,yy7}.
    Furthermore, the quantum entanglement \cite{yy8,yy9,yy10} and quantum discord \cite{yy11} have singularity in the region of QPTs.
    One of the most widely studied quantum systems is the so-called $XY$ chain in a transverse field ($XY$TF) which has interesting QPTs belonging to two different universality classes \cite{f7}: Ising transition and anisotropic transition.
    They divide the phase space of $XY$TF into three regions: $x-$direction ferromagnetic (FM$_{x}$), $y-$direction ferromagnetic (FM$_{y}$) and paramagnetic (PM) phases.
    Moreover, a gapless chiral phase is induced by antisymmetric Dzyaloshinskii and Moriya (DM) interaction \cite{d5}.
    And another system which has an analogous gapless chiral phase is the $XY$TF with $XZY-YZX$ type of three-site interaction \cite{bc3}.

    On the other hand, it is found that there is a symmetric helical interaction accompanying with DM interaction in the seminal paper\cite{a2}. The symmetric helical interaction is one order of magnitude smaller than the DM interaction and thus is neglected.
    Then, Kaplan find this symmetric helical interaction between the local spins in the single-band Hubbard model with spin-orbit couplings (SOCs) \cite{e4}.
    Afterwards, Shekhtman, Entin-Wohlman, and Aharony find that the weak ferromagnetism of La$_{2}$CuO$_{4}$ can be explained by this non-negligible symmetric helical interaction \cite{e5}.
    Therefore, the symmetric helical interaction is called KSEA interaction for short \cite{e6,e7,e8,e9,e10,f1,f2,f3,f4,f5,f6}.
    By magnetization measurements, neutron diffraction, and inelastic neutron scattering experiments, KSEA interaction is present in Ba$_{2}$CuGe$_{2}$O$_{7}$ \cite{e7,e8,e9,e10,f1}, Yb$_{4}$As$_{3}$ \cite{f2}, K$_{2}$V$_{3}$O$_{8}$ \cite{f3}, and La$_{2}$CuO$_{4}$ \cite{f4}.
    Meanwhile, the effects of KSEA interaction on the ground-state properties of spin$-$Peierls system \cite{f5,f6} have been studied.
    By regulating the topological insulator \cite{h7} and the system in the presence of Rashba spin-orbit coupling (RSOC) \cite{h9} out of equilibrium, they can synthesize and control the indirect KSEA interaction in two local spins.
    Therefore, it is interesting to study the effects of KSEA interaction on QPTs.
    In this paper, we study the effects of KSEA interaction on the ground-state properties of three kinds of chains: $XY$TF, $XY$TF with DM interaction, $XY$TF with the $XZY - YZX$ type of three-site interaction.

    This paper is organized as follows. In Section~\uppercase\expandafter{\romannumeral2}, we introduce the Hamiltonians of the three kinds of chains. In Section~\uppercase\expandafter{\romannumeral3}, The phase diagrams are given by means of correlation functions and chiral order parameters. A brief conclusion is given in Section~\uppercase\expandafter{\romannumeral4}.

\section{Hamiltonians}
    The Hamiltonians of the $XY$TF, $XY$TF with DM interaction and $XY$TF with $XZY - YZX$ type of three-site interaction are

    \begin{equation}
    \begin{split}
    H_{1}^{0}=&-\sum_{n=1}^{N}\{\frac{J}{2}[(1+\gamma)\sigma_{n}^{x}\sigma_{n+1}^{x}+
    (1-\gamma)\sigma_{n}^{y}\sigma_{n+1}^{y}]+h\sigma_{n}^{z}\},
    \label{h0}
    \end{split}\nonumber
    \end{equation}
    \begin{equation}
    \begin{split}
    H_{2}^{0}=&-\sum_{n=1}^{N}\{\frac{J}{2}[(1+\gamma)\sigma_{n}^{x}\sigma_{n+1}^{x}+
    (1-\gamma)\sigma_{n}^{y}\sigma_{n+1}^{y}]+h\sigma_{n}^{z}\}\\
    &-\sum_{n=1}^{N}D(\sigma_{n}^{x}\sigma_{n+1}^{y}-\sigma_{n}^{y}\sigma_{n+1}^{x})
    \label{h01}
    \end{split}\nonumber
    \end{equation}
    and
    \begin{equation}
    \begin{split}
    H_{3}^{0}=&-\sum_{n=1}^{N}\{\frac{J}{2}[(1+\gamma)\sigma_{n}^{x}\sigma_{n+1}^{x}+
    (1-\gamma)\sigma_{n}^{y}\sigma_{n+1}^{y}]+h\sigma_{n}^{z}\}\\
    &-\sum_{n=1}^{N}F(\sigma_{n-1}^{x}\sigma_{n}^{z}\sigma_{n+1}^{y}-\sigma_{n-1}^{y}\sigma_{n}^{z}\sigma_{n+1}^{x}),
    \label{h02}
    \end{split}\nonumber
    \end{equation}
    respectively.
    Here, the $\sigma_{n}^{x,y,z}$ are the Pauli matrices, $J$ is the nearest neighbor interactions (without loss of generality, we take $J=1$ in this paper), $h$ is a uniform external transverse field, $-1\leq\gamma\leq1$ is a parameter characterizing the degree of anisotropy of the interactions in the $xy$ plane, $D$ and $F$ are the coefficient of DM and $XZY-YZX$ type of three-site interaction, respectively.

    In generally, the KSEA interaction can be written as \cite{f6}
    \begin{equation}
    \begin{split}
    H_{\rm KSEA}=\sum_{n=1}^{N}\left(\begin{array}{ccc}\sigma_{n}^{x}&\sigma_{n}^{y}&\sigma_{n}^{z}\end{array}\right)
    \left(\begin{array}{ccc}0&a_{12}&a_{13}\\a_{12}&0&a_{23}\\a_{13}&a_{23}&0
    \end{array}\right)
    \left(\begin{array}{c}\sigma_{n+1}^{x}\\ \sigma_{n+1}^{y}\\ \sigma_{n+1}^{z}\end{array}\right)\nonumber.
    \end{split}
    \end{equation}


    Similar to reference \cite{f6}, we consider a special kind of KSEA interaction: $a_{12}=K$, $a_{13}=0$ and $a_{23}=0$ without loss of generality. And we study the ground state properties of three kinds of chains whose Hamiltonians are

    \begin{equation}
    \begin{split}
    H_{l}=-\sum_{n=1}^{N}K(\sigma_{n}^{x}\sigma_{n+1}^{y}+\sigma_{n}^{y}\sigma_{n+1}^{x})+H_{l}^{0},
    \label{eqb}
    \end{split}
    \end{equation}
    where $l=1,2$ or $3$.

    By using the Jordan-Wigner, the Hamiltonians (\ref{eqb}) can be written by spinless fermion operator, which yield
    \begin{equation}
    \begin{split}
    H_{l}=&-\sum_{n=1}^{N}[A_{l}c_{n}^{\dagger}c_{n+1}+B_{l}c_{n}^{\dagger}c_{n+1}^{\dagger}+
    C_{l}c_{n-1}^{\dagger}c_{n+1}+\\
    &h(c_{n}^{\dagger}c_{n}-\frac{1}{2})]+\rm{H.C.},
    \label{eqaa}
    \end{split}
    \end{equation}
    where $A_{1}=A_{3}=1$, $A_{2}=1+2D\rm{i}$, $B_{1}=B_{2}=B_{3}=\gamma-2K\rm{i}$, $C_{1}=C_{2}=0$ and $C_{3}=-2F\rm{i}$, respectively.
    Through Fourier transforms of the fermionic operators $c_{n}=\Sigma_{k}c_{k}e^{-{\rm i}kn}/\sqrt{N}$ and the Bogoliubov transformation $c_{k}=u_{k}\eta_{k}+v_{-k}^{\star}\eta_{k}^{\dagger}$ (see $u_{k}$ and $v_{k}$ in the Appendix A), Hamiltonians (\ref{eqaa}) can be written in the momentum space and be reduced to the diagonal form \cite{f7}

    \begin{align}
    H_{l}=\sum_{k=-\pi}^{\pi}\varepsilon_{l,k}(\eta_{k}^{\dag}\eta_{k}-1/2),
    \label{eqd}
    \end{align}
    where
    \begin{align}
    \varepsilon_{1,k}=\sqrt{({\rm cos}k+h)^{2}+(\gamma^{2}+4K^{2}){\rm sin}^{2}k},
    \label{eqe}
    \end{align}
    \begin{align}
    \varepsilon_{2,k}=-2D{\rm sin}k\pm\sqrt{({\rm cos}k+h)^{2}+(\gamma^{2}+4K^{2}){\rm sin}^{2}k}
    \label{eq4}
    \end{align}
    and
    \begin{align}
    \varepsilon_{3,k}=2F{\rm sin}(2k)\pm\sqrt{({\rm cos}k+h)^{2}+(\gamma^{2}+4K^{2}){\rm sin}^{2}k},
    \label{eq7}
    \end{align}
    respectively.

\section{phase diagrams}
\subsection{$XY$TF}
    In the light of the symmetry of Hamiltonian (\ref{eqb}), here we consider the case of $-1\leq\gamma\leq1$, $K$, $D$ and $F\geq0$.
    In the $XY$TF with KESA interaction, we can use Lee$-$Yang zeros method \cite{bc1,bc2} to figure out the critical points. The zeros of $\varepsilon_{1,k}$ in the complex $h$ plane are
    \begin{figure}
      \centering
      \includegraphics[width=.45\textwidth]{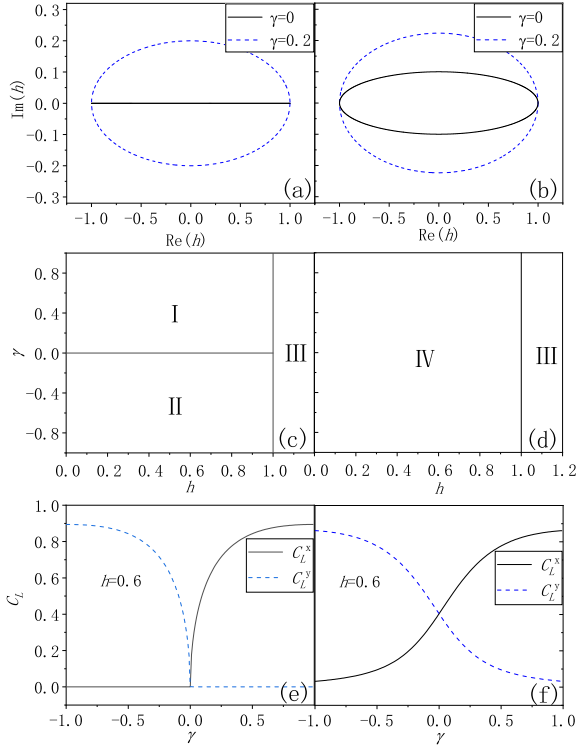}
      \caption{(a) and (b) are the solutions of $\varepsilon(h) = 0$ in the complex $h$ plane. (c) and (d) are the phase diagrams. (e) and (f) are the correlation functions $C_{L}^{x}$ as a function of $\gamma$. (a), (c) and (e) $K=0$. (b), (d) and (f) $K=0.2$, respectively.}
      \label{fig1}
    \end{figure}

    \begin{align}
    h=-{\rm cos}(k)+{\rm i}\sqrt{4K^{2}+\gamma^{2}}{\rm sin}(k).
    \label{eq2}
    \end{align}

    The diagrams of zeros are given in the Fig. \ref{fig1}(a) and (b) for $K=0$ and $K\neq0$, respectively.
    Without KSEA interaction ($K=0$), the zeros are located at the line segment of $-1\leq h\leq1$ when $\gamma=0$ [see the solid line in Fig. \ref{fig1}(a)].
    It describe the phase transition point of anisotropic transition.
    When $\gamma\ne0$, the zeros lie on an ellipse which intersects the real axis at $h=\pm1$ [see the dashed line in Fig. \ref{fig1}(a)].
    It corresponds to Ising transition.
    The corresponding phase diagram are showed in the Fig. \ref{fig1}(c).
    While $K\neq0$, the zeros are located at a ellipse which intersects the real axis at $h=\pm1$ even $\gamma=0$ [see Fig. \ref{fig1}(b)].
    Different from that of $K=0$, it means that the anisotropic transition disappears because of the KSEA interaction [see Fig. \ref{fig1}(d)].

    To understand the properties of the phases we calculated the correlation functions.
    The correlation functions are defined as
    \begin{align}
    C_{L}^{x}=\langle\sigma_{1}^{x}\sigma_{L}^{x}\rangle=\langle B_{1}A_{2}B_{2}A_{3}\ldots B_{L-1}A_{L}\rangle,\nonumber
    \end{align}
    \begin{align}
    C_{L}^{y}=\langle\sigma_{1}^{x}\sigma_{L}^{y}\rangle=(-1)^{L-1}\langle A_{1}B_{2}A_{2}B_{3}\ldots A_{L-1}B_{L}\rangle\nonumber
    \end{align}
    with $A_{n}=c_{n}^{\dag}+c_{n}$ and $B_{n}=c_{n}^{\dag}-c_{n}$. The $C_{L}^{x}$ and $C_{L}^{y}$ can be calculated by using the Wick theorem. The basic contractions of Wick theorem are

    \begin{equation}
     \begin{split}
     G_{lm}&=\langle B_{l}A_{m}\rangle=-\langle A_{m}B_{l}\rangle\\
     &=\frac{1}{N}\sum_{k=-\pi}^{\pi} \theta(\varepsilon_{k}) e^{-{\rm i}k(l-m)}[(v_{k}-u_{k})(u_{k}^{\star}+v_{k}^{\star})]\\
     &+\frac{1}{N}\sum_{k=-\pi}^{\pi} \theta(-\varepsilon_{k}) e^{{\rm i}k(l-m)}[(v_{k}+u_{k})(u_{k}^{\star}-v_{k}^{\star})],
     \end{split}
     \label{eqn}
     \end{equation}
     \begin{equation}
     \begin{split}
     Q_{lm}&=\langle A_{l}A_{m}\rangle\\
     &=\frac{1}{N}\sum_{k=-\pi}^{\pi}e^{-{\rm i} \theta(\varepsilon_{k}) k(l-m)}[(v_{k}+u_{k})(u_{k}^{\star}+v_{k}^{\star})],
     \end{split}
     \label{eqo}
     \end{equation}
     \begin{equation}
     \begin{split}
     P_{lm}&=\langle B_{l}B_{m}\rangle\\
     &=\frac{1}{N}\sum_{k=-\pi}^{\pi}e^{-{\rm i} \theta(\varepsilon_{k}) k(l-m)}[(v_{k}-u_{k})(u_{k}^{\star}-v_{k}^{\star})],
     \end{split}
     \label{eqp}
     \end{equation}
     where
    \begin{equation}
    \theta(x)=
    \left\{
                 \begin{array}{lr}
                 1, &  x\ge0\\
                 0, &  x<0.
                 \end{array}
    \right.
    \end{equation}

    If $K=0$, $Q_{lm}$ and $P_{lm}$ always zero. The correlation functions can be simply calculated by $G_{lm}$ (See Appendix B).
    However, when $K\neq0$, the form of $u_{k}$ and $v_{k}$ are different from those of the $XY$TF without KSEA interaction so that $Q_{lm}$ and $P_{lm}$ nonzero anymore. We must use Pfaffian to figure out the correlation functions (See Appendix B).

    The results of correlation functions are given in Fig. \ref{fig1}(e) and (f) with $K=0$ and $K\neq0$, respectively.
    They are not differentiable at $\gamma=0$ when $K=0$ (see it in Fig. \ref{fig1}(e)) but differentiable at $\gamma=0$ when $K\neq0$ (see it in Fig. \ref{fig1}(f)).
    According to the correlation functions we study the properties of the phases.
    In region \uppercase\expandafter{\romannumeral1} in Fig. \ref{fig1}(c), the $C_{L}^{x}\neq0$ and $C_{L}^{y}=0$ corresponding to FM$_{x}$ phase. While the $C_{L}^{x}=0$ and $C_{L}^{y}\neq0$ correspond to FM$_{y}$ phase in the region \uppercase\expandafter{\romannumeral2} in Fig. \ref{fig1}(c).
    The KSEA interaction has no effect on the properties of region \uppercase\expandafter{\romannumeral3} in Fig. \ref{fig1}(c) which corresponds to PM phase.
    There is a new region \uppercase\expandafter{\romannumeral4} in the phase space in Fig. \ref{fig1}(d).
    In this region, the $C_{L}^{x}\neq0$ and $C_{L}^{y}\neq0$ correspond to a general ferromagnetic phase which is in the $x-$direction and $y-$direction (FM$_{xy}$ phase).
\subsection{$XY$TF with DM interaction}
    \begin{figure}
      \centering
      \includegraphics[width=.45\textwidth]{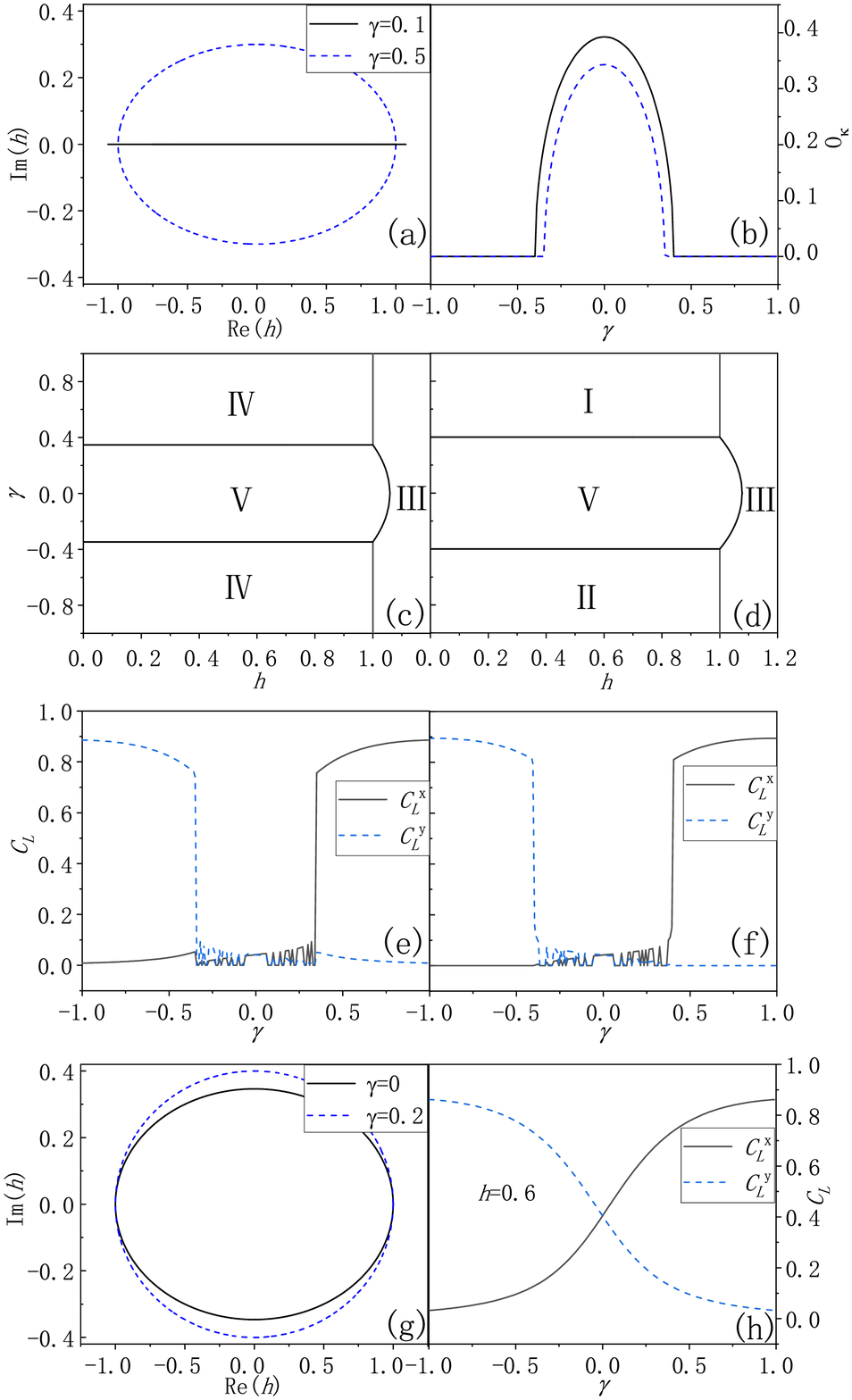}
      \caption{(a) and (g) are the solutions of $\varepsilon(h) = 0$ in the complex $h$ plane. (b) is the chiral order parameter as a function of $\gamma$. The solid and dashed lines correspond to the $K=0$ and $K=0.1$ with the same $D=0.2$, respectively. (c) and (d) are the phase diagrams. (e), (f) and (h) are the correlation functions as a function of $\gamma$. (a), (c) and (e) $K=0.1$, $D=0.2$, (d) and (f) $K=0$, $D=0.2$. (g) and (h) $K>D=0.2$.}
      \label{fig2}
    \end{figure}
    Similarly, we also use the Lee$-$Yang zeros method to calculate the critical points on the complex $h$ plane for KESA interaction added on $XY$TF with DM interaction. The zeros in the complex $h$ plane can be written as
    \begin{align}
    h=-{\rm cos}(k)+\sqrt{4(D^{2}-K^{2})-\gamma^{2}}{\rm sin}(k).
    \label{eq5}
    \end{align}
    There are two possible scenarios.

    (\uppercase{\romannumeral1}) $K<D$.
    The zeros with different $\gamma$ in the complex $h$ plane are given in Fig. \ref{fig2}(a).
    It is easy to check that the zeros, which lie on an ellipse, are cutting the real axis at $h = \pm1$ in the complex $h$ plane if $|\gamma| >2\sqrt{D^{2}-K^{2}}$ [see the dashed line in Fig. \ref{fig2}(a)].
    If $|\gamma| \leq2\sqrt{D^{2}-K^{2}}$, from  (\ref{eq5}) we can see that the zeros lie on the real axis in the interval $|h|\leq\sqrt{4(D^{2}-K^{2})-\gamma^{2}+1}$.
    It correspond to the solid line in Fig. \ref{fig2}(a).
    From the above formula, the region of chiral phase in the parameter space decreases as $K$ increasing.

    In Fig. \ref{fig2}(c) and (e), phase diagram and correlation functions are shown respectively for $K<D$.
    In the two regions \uppercase\expandafter{\romannumeral4} in Fig. \ref{fig2}(c), the $C_{L}^{x}\neq0$ and $C_{L}^{y}\neq0$ corresponding to FM$_{xy}$ phase.
    The direction of the ferromagnetism is inclined to $x-$direction when $\gamma>0$.
    For comparison, the phase diagram and correlation functions corresponding to the coefficients $K=0$ and $D=0.2$ are given in Fig. \ref{fig2}(d) and (f), respectively.
    In region \uppercase\expandafter{\romannumeral1} in Fig. \ref{fig2}(d), the $C_{L}^{x}\neq0$ and $C_{L}^{y}=0$ corresponding to FM$_{x}$ phase. The $C_{L}^{x}=0$ and $C_{L}^{y}\neq0$ corresponding to FM$_{y}$ phase in the region \uppercase\expandafter{\romannumeral2} in Fig. \ref{fig2}(d).

    However, the correlation functions are not a good order parameters in the gapless chiral phase so that we calculated the chiral order parameter in the $z$-direction \cite{d5}.

     \begin{align}
     O_{\kappa}=\frac{1}{N}\sum_{n=1}^{N}\langle \boldsymbol{k} \cdot \boldsymbol{\sigma}_{n} \times \boldsymbol{\sigma}_{n+1}\rangle,
     \end{align}
     where $\boldsymbol{k}$ is the unit vector in $z$-direction.

     We can obtain the chiral order parameter
     \begin{align}
     O_{\kappa}=-\frac{4}{N}\sum_{k} {\rm sin}k\Theta(\varepsilon_{2,k}),
     \end{align}
     where
     \begin{align}
     \Theta(x)=\{\begin{array}{c}1,\ x\geq0,\\ 0,\ x<0.\end{array}
     \end{align}

     Obviously, $O_{\kappa}=0$ in regions \uppercase\expandafter{\romannumeral1}, \uppercase\expandafter{\romannumeral2} and \uppercase\expandafter{\romannumeral3} because $\varepsilon_{k}$ are always positive. The solid and dashed lines correspond to the coefficients $K=0$ and $K=0.1$ with a same $D=0.2$, respectively (see Fig. \ref{fig2}(b)).
     Similarly, the region of chiral phase in the parameter space decreases as $K$ increasing.

    (\uppercase{\romannumeral2}) $K>D$. Equation (\ref{eq5}) can be transformed to $h=$ $-{\rm cos}(k)+{\rm i}\sqrt{\gamma^{2}-4(D^{2}-K^{2})}{\rm sin}(k)$.
    The zeros are given in Fig. \ref{fig2}(g) for $\gamma=0$ and $\gamma=0.2$, respectively.
    It is found that the zeros are located at different ellipses with different $\gamma$ intersecting the real axis at $h=\pm1$.
    It is similar to that in the $XY$TF with KSEA interaction in Fig. \ref{fig1}(b).

    In Fig. \ref{fig2}(h) we give the correlation functions when $K>D$. We find that the correlation functions are differentiable about $-1\leq\gamma\leq1$ which is similar as Fig. \ref{fig1}(f). The $C_{L}^{x}$ and $C_{L}^{y}$ are nonzero in the whole region of $\gamma$. So the phase diagram is the same as Fig. \ref{fig1}(d) when $K>D$.

\subsection{$XY$TF with $XZY-YZX$
type of three-site interaction}
    \begin{figure}
    \centering
    \includegraphics[width=.45\textwidth]{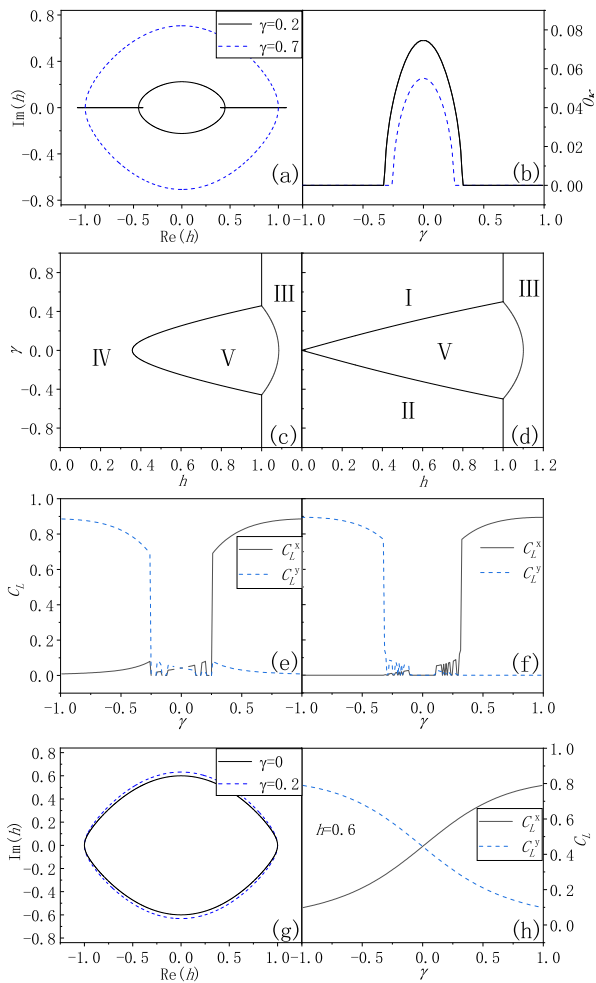}
    \caption{(a) and (g) are the solutions of $\varepsilon(h) = 0$ in the complex $h$ plane. (b) is the chiral order parameter as a function of $\gamma$. The solid and dashed lines correspond to the $K=0$ and $K=0.1$ with the same $F=0.125$, respectively. (c) and (d) are the phase diagrams. (e), (f) and (h) are the correlation functions as a function of $\gamma$. (a), (c) and (e) $K=0.1$, $F=0.125$, (d) and (f) $K=0$, $D=0.125$. (g) and (h) $K>D=0.125$.}
    \label{fig3}
    \end{figure}
    Similarly, the zeros in the complex $h$ plane can be written as
    \begin{align}
    h=-{\rm cos}(k)+\sqrt{16T^{2}{\rm cos}^{2}(k)-4K^{2}-\gamma^{2}}{\rm sin}(k).
    \label{eq8}
    \end{align}
    There are also two possible scenarios.

    (\uppercase{\romannumeral1}) $K<2F$.
    The zeros for $K=0.1$ and different $\gamma$ respectively in the complex $h$ plane are given in Fig. \ref{fig3}(a).
    It is easy to check that the zeros, which lie on an ellipse, are cutting the real axis at $h = \pm1$ in the complex $h$ plane if $\gamma>2\sqrt{(2F)^{2}-K^{2}}$ (see the dashed line in Fig. \ref{fig2}(a)).
    When $\gamma\leq2\sqrt{(2F)^{2}-K^{2}}$, the zeros are located on two line segments on the real $h$ axis [see the solid line in Fig. \ref{fig3}(a)].

    In Fig. \ref{fig3}(c) and (e), we give the phase diagram and correlation functions respectively for $K<2F$.
    In the regions \uppercase\expandafter{\romannumeral4} in Fig. \ref{fig3}(c), the $C_{L}^{x}\neq0$ and $C_{L}^{y}\neq0$ corresponding to FM$_{xy}$.
    We also find that the direction of the ferromagnetism is inclined to $x-$direction when $\gamma>0$.
    For comparison, the phase diagram and correlation functions corresponding to the coefficients $K=0$ and $F=0.125$ are shown in Fig. \ref{fig3}(d) and (f), respectively.
    In region \uppercase\expandafter{\romannumeral1} in Fig. \ref{fig3}(d)), the $C_{L}^{x}\neq0$ and $C_{L}^{y}=0$ corresponding to FM$_{x}$. The $C_{L}^{x}=0$ and $C_{L}^{y}\neq0$ corresponding to FM$_{y}$ in the region \uppercase\expandafter{\romannumeral2} in Fig. \ref{fig3}(d).

    To understand the properties of the phase \uppercase\expandafter{\romannumeral5} in Fig. \ref{fig3}(d), we calculate the chiral order parameters as follow

    \begin{equation}
     \begin{split}
     O_{\kappa}=\frac{1}{N}\sum_{n=1}^{N}\langle \boldsymbol{\sigma}_{n-1}\cdot(\boldsymbol{\sigma}_{n}\times\boldsymbol{\sigma}_{n+1})\rangle.
     \label{eq9}
     \end{split}
     \end{equation}

    After the calculation of $O_{\kappa}$ we can obtain
    \begin{equation}
     \begin{split}
     O_{\kappa}=&-\frac{4}{N}\sum_{k}[2(1-G_{i+1,i}-G_{i,i+1}){\rm cos}(k)\\
     &+(2G_{i,i}+G_{i+2,i}+G_{i,i+2})]\Theta(\varepsilon_{3,k}){\rm sin}(k),
     \label{eq10}
     \end{split}
     \end{equation}
    where the basic contractions $\langle G_{lm}\rangle$ are the same as Equation (\ref{eqn}) with a different form of $u_{k}$ and $v_{k}$ (See Appendix A). On the basis of translation invariance, all $G_{i,i+j}$ are the same with different $i$ and the same $j$. The chiral order parameters in region \uppercase\expandafter{\romannumeral1}, \uppercase\expandafter{\romannumeral2}, \uppercase\expandafter{\romannumeral3} and \uppercase\expandafter{\romannumeral4} are zero apparently and nonzero in the region \uppercase\expandafter{\romannumeral5}.
    The solid and dashed lines correspond to the coefficients $K=0$ and $K=0.1$ with a same $F=0.125$, respectively (see Fig. \ref{fig3}(b)).
    Similarly, the region of chiral phase in the parameter space decreases when $K$ increased.

    (\uppercase{\romannumeral2}) $K>2F$. Equation (\ref{eq8}) can be transformed to $h=$ $-{\rm cos}(k)+{\rm i}\sqrt{\gamma^{2}-4[4F^{2}{\rm cos}^{2}(k)-K^{2}]}{\rm sin}(k)$.
    The diagram of zeros is given in Fig. \ref{fig3}(g) with $\gamma=0$ and $\gamma=0.2$, respectively.
    It is found that the zeros are located at different ellipses with different $\gamma$ intersecting the real axis at $h=\pm1$.
    It is the same as the situation of $XY$TF with KSEA interaction in Fig. \ref{fig1}(b).

    In Fig. \ref{fig3}(h) we give the correlation functions when $K>D$. We find that the correlation functions are differentiable in the interval $[-1,1]$ which is similar as Fig. \ref{fig1}(f). The $C_{L}^{x}$ and $C_{L}^{y}$ are nonzero in the whole region of $\gamma$. So the phase diagram when $K>D$ is the same as Fig. \ref{fig1}(d).



\section{Conclusion}
    In summary, we have studied the effects of the KSEA interactions on the ground-state properties of the three kinds of spin chains which are $XY$TF, $XY$TF with DM interaction, $XY$TF with $XZY-YZX$ type of three-site interaction.
    In the first case, the zeros lying on the segment of real axis of the complex $h$ plane disappear when $K$ is nonzero. The correlation functions turn into differentiable at the point where $\gamma=0$. So the anisotropic transition disappear because of the KSEA interaction no matter how small it is.

    In the second case, if KSEA interaction is bigger than DM interaction, the system is similar to the $XY$TF with KSEA interaction.
    If KSEA interaction is smaller than DM interaction, the region of chiral phase in the parameter space decreases as $K$ increasing.
    The FM$_{x}$ and FM$_{y}$ phases of the $XY$TF with DM interaction turned into a FM$_{xy}$ phase when $K$ is nonzero.
    In the FM$_{xy}$ phase, the direction of the ferromagnetism is inclined to $x-$direction when $\gamma$ is greater than 0.

    In the third case, if KSEA interaction is bigger than twice $XZY-YZX$ type of three-site interaction, the system is similar to the $XY$TF with KSEA interaction too. If KSEA interaction is smaller than twice $XZY-YZX$ type of three-site interaction, the region of chiral phase in the parameter space decreases as $K$ increasing. The FM$_{x}$ and FM$_{y}$ phases of the $XY$TF with $XZY-YZX$ type of three-site interaction turned into a FM$_{xy}$ phase when $K$ is nonzero. In addition, the two phases are connected when $K$ is nonzero which is different from the previous chain.
    In the FM$_{xy}$ phase, the direction of the ferromagnetism is inclined to $x-$direction when $\gamma$ is greater than 0.
\section*{ACKNOWLEDGMENTS}
    The  work is supported by the National Natural Science Foundation of China (Grant Nos. 11575057 and 11975126).
\section{appendix A}
     In the Bogoliubov transformations, the $u_{k}$ and $v_{k}$ are different from the cases without KSEA interaction. They are
     \begin{align}
     u_{k}=\frac{(2K+\gamma {\rm i}){\rm sin}k}{\sqrt{(\varepsilon_{k}+{\rm cos}k+h)^{2}+(4K^{2}+\gamma^{2}){\rm sin}^{2}k}},\nonumber \\ v_{k}=\frac{\varepsilon_{k}+{\rm cos}k+h}{\sqrt{(\varepsilon_{k}+{\rm cos}k+h)^{2}
    +(4K^{2}+\gamma^{2}){\rm sin}^{2}k}},\nonumber
    \end{align}
    for KSEA interaction added on $XY$ chain,
    \begin{align}
    u_{k}=\frac{(2K+\gamma {\rm i}){\rm sin}k}{\sqrt{(\varepsilon_{k}+{\rm cos}k+h+2D{\rm sin}k)^{2}+(4K^{2}+\gamma^{2}){\rm sin}^{2}k}},\nonumber \\ v_{k}=\frac{\varepsilon_{k}+{\rm cos}k+h+2D{\rm sin}k}{\sqrt{(\varepsilon_{k}+{\rm cos}k+h+2D{\rm sin}k)^{2}
    +(4K^{2}+\gamma^{2}){\rm sin}^{2}k}},\nonumber
    \end{align}
    for KSEA interaction added on $XY$ chain with DM interaction and
    \begin{align}
    u_{k}=\frac{(2K+\gamma {\rm i}){\rm sin}k}{\sqrt{[\varepsilon_{k}+{\rm cos}k+h-2F{\rm sin}(2k)]^{2}+(4K^{2}+\gamma^{2}){\rm sin}^{2}k}},\nonumber \\ v_{k}=\frac{\varepsilon_{k}+{\rm cos}k+h-2F{\rm sin}(2k)}{\sqrt{[\varepsilon_{k}+{\rm cos}k+h-2F{\rm sin}(2k)]^{2}
    +(4K^{2}+\gamma^{2}){\rm sin}^{2}k}},\nonumber
    \end{align}
    for KSEA interaction added on $XY$ chain with $xzy-yzx$ type of three-site interaction.
\section{appendix B}
     If $K=0$ and all $\varepsilon_{i,k}\geq0$, the correlation functions can be written as
     \begin{align}
     C_{L}^{x}=\left|\begin{array}{cccc}G_{12}&G_{13}& \cdots &G_{1L}\\ \vdots&\vdots&\vdots&\vdots\\ \vdots &\vdots&\vdots&\vdots\\G_{L-1,2}&G_{L-1,3}& \cdots &G_{L-1,L}
     \end{array}\right|,\nonumber
     \end{align}
     \begin{align}
     C_{L}^{y}=\left|\begin{array}{cccc}G_{21}&G_{22}& \cdots &G_{2,L-1}\\ \vdots&\vdots&\vdots&\vdots\\ \vdots &\vdots&\vdots&\vdots\\G_{L1}&G_{L2}& \cdots &G_{L,L-1}
     \end{array}\right|.\nonumber
     \end{align}

     If $K\ne0$ or $\varepsilon_{i,k}\le0$ the correlation functions are \cite{h10}
     \begin{align}
     C_{L}^{x}=(F_{x})^{1/2},
     \label{eqq}
     \end{align}
     \begin{align}
     C_{L}^{y}=(F_{y})^{1/2},
     \label{eqr}
     \end{align}
     where
     \begin{align}
     F_{x}=\left|\begin{array}{ccccccc}0&G_{12}&P_{12}&G_{13}&P_{13}& \cdots &G_{1n}\\ -G_{12}&0&-G_{22}&Q_{23}&-G_{32}& \cdots &Q_{2n}\\ -P_{12}&G_{22}&0&G_{23}&P_{23}& \cdots &G_{2n}
     \\ \vdots&\vdots&\vdots&\vdots&\vdots&\vdots&\vdots\\ -G_{1n}&-Q_{2n}&\cdots&\cdots&\cdots&\cdots&0
     \end{array}\right|,\nonumber
     \end{align}
     \begin{align}
     F_{y}=\left|\begin{array}{ccccccc}0&-G_{21}&Q_{12}&-G_{31}&Q_{13}& \cdots &-G_{n1}\\ G_{21}&0&G_{22}&P_{23}&G_{23}& \cdots &P_{2n}\\ -Q_{12}&-G_{22}&0&-G_{32}&Q_{23}& \cdots &-G_{n2}
     \\ \vdots&\vdots&\vdots&\vdots&\vdots&\vdots&\vdots\\ G_{n1}&-P_{2n}&\cdots&\cdots&\cdots&\cdots&0
     \end{array}\right|.\nonumber
     \end{align}


\end{document}